\documentclass[trackchanges]{aastex701}
\usepackage{color}
\usepackage{graphicx}
\usepackage{epstopdf}
\usepackage{latexsym}
\usepackage{threeparttable}
\usepackage{rotating}
\usepackage{CJK}

\begin{document}
\begin{CJK*}{UTF8}{gbsn}

\title{The identification of new Herbig Ae/Be stars from LAMOST DR7}

\correspondingauthor{Jiaming, Liu; Wenyuan, Cui}
\email{jmliu@hebtu.edu.cn; cuiwenyuan@hebtu.edu.cn;}

\author[0009-0001-2026-368X]{Jialin Liu (刘佳琳)}
\email{liujialin241@mails.ucas.ac.cn}
\affiliation{College of Physics, Hebei Normal University, Shijiazhuang 050024, People's Republic of China}
\affiliation{National Astronomical Observatories, Chinese Academy of Sciences, Beijing 100101, People's Republic of China}
\author[0000-0002-0786-7307]{Jiaming Liu (刘佳明)}
\email{jmliu@hebtu.edu.cn}
\affiliation{College of Physics, Hebei Normal University, Shijiazhuang 050024, People's Republic of China}
\affiliation{Guo Shoujing Institute for Astronomy, Hebei normal University, Shijiazhuang 050024, People's Republic of China}
\author{Jiya Yao (姚季雅)}
\email{jiya.yao@studio.unibo.it}
\affiliation{Department of Physics and Astronomy ``Augusto Righ'', University of Bologna, via Gobetti 93/2, I-40129 Bologna, Italy}
\author{Zhenghao Cheng (程正昊)}
\email{zhcheng@hebtu.edu.cn}
\affiliation{College of Physics, Hebei Normal University, Shijiazhuang 050024, People's Republic of China}
\author[0009-0005-5413-7664]{Qingyue Qu (曲卿乐)}
\email{Quqy@bao.ac.cn}
\affiliation{College of Physics, Hebei Normal University, Shijiazhuang 050024, People's Republic of China}
\affiliation{National Astronomical Observatories, Chinese Academy of Sciences, Beijing 100101, People's Republic of China}
\author[0000-0001-5314-2924]{Zhicun Liu (柳志存)}
\email{liuzhicun@hebtu.edu.cn}
\affiliation{College of Physics, Hebei Normal University, Shijiazhuang 050024, People's Republic of China}
\affiliation{Guo Shoujing Institute for Astronomy, Hebei normal University, Shijiazhuang 050024, People's Republic of China}
\author[0000-0003-1359-9908]{Wenyuan Cui (崔文元)}
\email{cuiwenyuan@hebtu.edu.cn}
\affiliation{College of Physics, Hebei Normal University, Shijiazhuang 050024, People's Republic of China}
\affiliation{Guo Shoujing Institute for Astronomy, Hebei normal University, Shijiazhuang 050024, People's Republic of China}
\author[0000-0001-8060-1321]{Min Fang (房敏)}
\email{mfang@pmo.ac.cn}
\affiliation{Purple Mountain Observatory, Chinese Academy of Sciences, 10 Yuanhua Road, Nanjing 210023, People’s Republic of China}

\begin{abstract}
Herbig Ae/Be stars (HAeBes) are critical tracers of intermediate- and high-mass star formation, yet their census remains incomplete compared to low-mass young stellar objects like T-Tauri stars. To expand the known population, we systematically searched for HAeBes in LAMOST DR7 low-resolution spectra. Following Sun et al., we applied Uniform Manifold Approximation and Projection (UMAP) for dimensionality reduction and Support Vector Machine (SVM) classification, identifying $\sim$240,000 spectra with potential H$\alpha$ emission. After removing contaminants (non-stellar objects, extragalactic sources, CVs, and Algol systems) and restricting to B/A-type stars, we obtained 1,835 candidates through 2MASS/WISE visual inspection. Spectral energy distribution analysis confirmed 143 sources with infrared excess ($J$-band or longer wavelengths), including 92 known HAeBes. From the remaining 51 candidates, we classified 26 with strong infrared excess as new HAeBes. Color-index analysis of confirmed HAeBes and classical Ae/Be stars (CAeBes) revealed that the $(K-W1)_0$ vs. $(W2-W3)_0$ diagram effectively separates these populations: CAeBes predominantly occupy $(K-W1)_0 \leq 0.5$ and $(W2-W3)_0 \leq 1.1$, while other regions trace transition disks ($(K-W1)_0 < 0.5$ and $(W2-W3)_0 > 1.1$), globally depleted disks ($(K-W1)_0 > 0.5$ and $(W2-W3)_0 < 1.1$), and Class I/Flat/II HAeBes ($(K-W1)_0 > 0.5$ and $(W2-W3)_0 > 1.1$). More importantly, the HAeBes exhibit a clear evolutionary gradient on this diagram, with those in the Class III, Class II, Flat-SED, and Class I evolutionary stages being effectively distinguished by concentric ellipses that are roughly centered at (0,0) with semi-major axes of $a$=1.5, $a$=3.0, and $a$=4.0, and a semi-major to semi-minor axis ratio of 1.6:1.
\end{abstract}



\section{Introduction} \label{sec:intro}
%Strong emission lines can be caused by stellar wind and envelopes, even accretion flows of revealing \citep{Zhang_2022}.
Emission-line stars are characterized by prominent emission lines in their optical spectra, such as hydrogen lines and metallic lines. The most prominent feature of emission-line stars is the H$\alpha$ line, which peaks at approximately 6562.8~\AA. Based on their evolutionary status, \cite{kogure2007astrophysics} classified emission-line stars into four categories: early-type stars (e.g., Wolf--Rayet stars, Of-type stars, Oe/Be/Ae stars, and luminous blue variables), late-type stars (e.g., dMe stars, Mira variables, flare stars, red giants, and post-AGB stars), close binaries, and pre-main-sequence stars (PMSs). Among these, Herbig Ae/Be stars (HAeBes), which are early-type and typically high-mass objects \citep{Zha2022}, are particularly important for studying the evolution---especially the early evolutionary stages---of massive stars.
%Among them, Herbig Ae/Be (HAeBes) and classical Ae/Be stars (CAeBes) are the most ready identified and large in number, since the other types (WRs, LBVs, etc) are more rarely detected.

HAeBes are typical pre-main-sequence (PMS) stars that are still in the stage of contracting, with masses between 2 and 8\,$M_{\odot}$. The initial identification and definition of this stellar class were made by \cite{herbig1960spectra} in 1960 to illustrate the intermediate-mass counterpart of T-Tauri stars (low-mass PMS stars). Subsequent investigations by \cite{strom1972nature} revealed that HAeBes are typically surrounded by circumstellar materials with a disk-like structure. Hence, HAeBes usually show obvious H$\alpha$ emission line in the optical, excess emission in the infrared bands, as well as remarkable photometric variabilities. These characteristics are used as criteria for the identification of HAeBe stars. Since then, the sample set has been largely extended by follow-up works \citep{finkenzeller1984herbig,Mal1998,Her2004}.

As a comparison, the term Classical Ae/Be stars (CAeBes) is used to distinguish them from HAeBes and Algol-type systems \citep{Porter_2003}. They are classical A- or B-type stars that exhibit emission lines in the optical range, such as Balmer lines, singly-ionized metal lines, and neutral helium lines \citep{kogure2007astrophysics}. This similarity makes it difficult to distinguish HAeBes from CAeBes. \cite{rivinius2013classical} examined the identified CBes and concluded that they are rapidly rotating main-sequence B stars surrounded by outwardly diffusing gaseous (dust-free) Keplerian disks. The infrared excess of CAeBes originates from this hot ionized gas. Consequently, the IR excess of CAeBes is much smaller than that of HAeBes, especially at longer wavelengths \citep{finkenzeller1984herbig}, because the infrared excess emission of HAeBes is mainly due to thermal emission from dust in the circumstellar disk. This difference is often used as a strong criterion for distinguishing CAeBes from HAeBes \citep{gehrz1974infrared,cote1987iras,zhang2006infrared,hou2016catalog,vioque2018gaia,bhattacharyya2021identification,Zha2022}. Based on this theory, \cite{hou2016catalog} argued that the $(H-K)_{0}$ vs.\ $(K-W1)_{0}$ color--color diagram is an effective tool for separating CAeBes and HAeBes, as HAeBes typically occupy the region where $(H-K)_{0}>0.4$ and $(K-W1)_{0}>0.8$, whereas CAeBes appear in the region where $(H-K)_{0}<0.2$ and $(K-W1)_{0}<0.5$.

By far, the magnificent datasets of survey telescopes have strongly favored the identification of HAeBes, and studies based on these data have greatly improved our understanding of them. \cite{vioque2022identification} revisited the HAeBe candidates of \citet{Vio2020} and confirmed 128 stars on the basis of the HR diagram, infrared excess, the presence of emission lines, the profiles of the H$\alpha$ line, and the mass-accretion rates derived therefrom. Using optical spectra from LAMOST and infrared images from WISE, \citet{Zha2022} identified 62 new HAeBes through H$\alpha$ emission and infrared colour excess. They analysed the H$\alpha$ line and divided the spectra of HAeBes into ten types according to the line profiles. At present, about 500 HAeBes are known, but this sample is still too small to support statistically robust investigations of their properties. In view of this, we have carried out a census of HAeBes in the low-resolution spectra of LAMOST DR7. The data selection is described in Section~\ref{sec:Data selection}, the method in Section~\ref{sec:Method}, the results in Section~\ref{sec:result}, the discussion in Section~\ref{sec:Dicussion}, and a summary in Section~\ref{sec:Conclusion}.

\section{DATA}
\label{sec:Data selection}

\subsection{Spectra}
\label{subsec:Spectra}
The basic dataset of this work is the recently released low-resolution spectra from LAMOST (the Large Sky Area Multi-Object Fiber Spectroscopic Telescope).  LAMOST is a 4\,m quasi-meridian Schmidt telescope.  The primary step of this study is to search the LAMOST spectral database for as many H$\alpha$ emission-line stars as possible. Therefore, we utilise the complete low-resolution spectral catalogue of LAMOST DR7, which comprises 10,431,197 spectra covering the wavelength range 3690--9100~\AA\ at a resolving power of $R\approx1800$.

\subsection{Photometry}
\label{subsec:Photometry}
Since the color excess (or the infrared excess emission) that due to the thermal emission of dust in the circumstellar disk is also a prominent feature of HAeBes, thus the photometric data of the Wide-field Infrared Survey Explorer (WISE; \citealt{Wright_2010}) that cover the near- and mid-infrared bands, i.e. the $ W1$ (3.4$\rm \mu m$), $ W2$ (4.6$\rm \mu m$), $W3$ (12$\rm \mu m$) and $W4$ (22$\rm \mu m$) are considered in this work. And the near-infrared bands $J$, $H$, $K$ of the Two Micron All-Sky Survey (2MASS; \citealt{Skr2006}) were also adopted for same purpose. To ensure reliability, only photometric measurements with a quality flag of ‘B’ or better are retained.

\section{Identification of New Herbig Stars}\label{sec:Method}

\subsection{Stars with H$\alpha$ Emission}

The H$\alpha$ emission line is a strong indicator of emission-line stars and is regarded as a fundamental property of HAeBe stars. Therefore, when searching for HAeBes, our primary criterion is to identify stars that show H$\alpha$ emission in their optical spectra. To achieve this goal, following the approach of \cite{Sun2021} in identifying H$\alpha$ emission-line stars, we adopt the machine-learning algorithms UMAP (Uniform Manifold Approximation and Projection) and SVM (Support Vector Machine).

\begin{figure}
%\begin{interactive}{js}{interactive.tar.gz}
\plotone{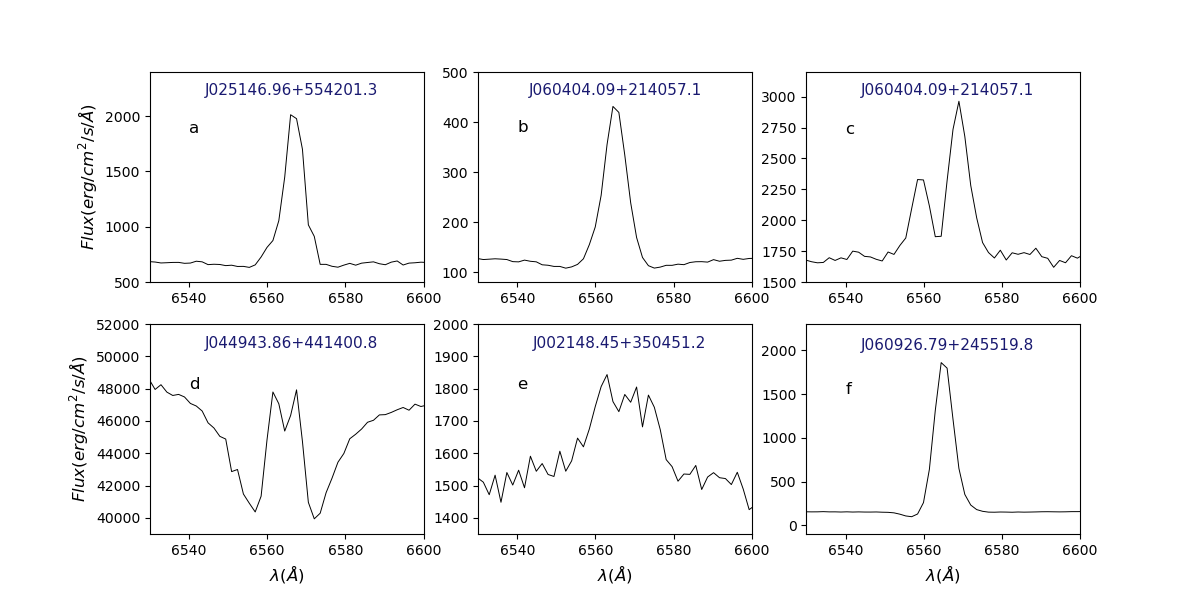}
%\end{interactive}
\centering
\caption{%Brief descriptions of each image are as follows:
Six morphological classifications of H$\alpha$ emission line profiles:
(a) Single-peak emission profile
(b) Single-peak emission superimposed on an absorption feature
(c) Double-peak emission profile
(d) Double-peak emission superimposed on an absorption feature
(e) Low signal-to-noise ratio (SNR) profile where the emission line remains detectable
(f) P Cygni profile (characterized by both emission and blueshifted absorption components)
}
\label{f1}
\end{figure}

UMAP is a dimensionality-reduction technique introduced by \cite{mcinnes2018umap}. Its core idea is to derive a low-dimensional representation of a large dataset based on the mathematical frameworks of algebraic topology and Riemannian geometry. Compared with the commonly used PCA (Principal Component Analysis), UMAP can provide clearer classification boundaries in parameter space and offers superior runtime performance, especially when handling high-dimensional, massive datasets \citep{Sun2021}. SVM is a supervised learning algorithm widely used for classification tasks. In this work, UMAP is first applied to the LAMOST spectral data to obtain a reduced-dimensional dataset, after which SVM is employed to select spectra with prominent H$\alpha$ emission profiles. Specifically, the procedure consists of three steps:

$\bullet$ Building the training sample. Ten catalogues of H$\alpha$ emission-line stars \citep{downes1993catalog,downes1997catalog,downes2001catalog,jiang2013data,szkody2011cataclysmic,breedt20141000,coppejans2016statistical,han2018cataclysmic,ritter2003catalogue,hou2020spectroscopically} were compiled. We cross-matched these catalogues with LAMOST DR7 spectra using a $2^{\prime\prime}$ search radius and visually inspected the corresponding spectra. In total, 400 stars showing clear H$\alpha$ emission were retained; the morphologies of their H$\alpha$ lines can generally be classified into six typical profiles (see Figure~\ref{f1}), reflecting different physical conditions in their circumstellar material. As a control sample, 1000 spectra without H$\alpha$ emission were randomly selected from LAMOST. Assigning label ``1'' to the 400 emission-line spectra and label ``0'' to the control sample yielded a training set of 1400 spectra.

$\bullet$ The machine-learning procedure. A wavelength window of 6530--6600~\AA\ was extracted from each spectrum to minimise contamination from other lines. After reducing the dataset to three dimensions with UMAP, we applied SVM to determine the optimal decision boundary for identifying H$\alpha$ emission stars. To ensure robustness, five-fold cross-validation was performed. With the parameters $n_{\mathrm{neighbors}}=27$ and $\min_{\mathrm{dist}}=0.02$, the average results across the folds are summarized as follows: 394 of the 400 (98.5\%) emission-line spectra were correctly recovered (true positives), while 6 (1.5\%) were missed (false negatives). From the 1000 control spectra, 887 (88.7\%) were correctly identified (true negatives), and 113 (11.3\%) were misclassified as emission-line sources (false positives). The overall classification accuracy was 91.5\%. 

$\bullet$ Application to LAMOST DR7. Using the above algorithm and parameters, we identified 241,156 candidate spectra with H$\alpha$ emission from the LAMOST DR7 dataset.

\subsection{Contaminant Removal}\label{sec:identification}

As demonstrated by \cite{kogure2007astrophysics}, more than a dozen astrophysical objects exhibit H$\alpha$ emission in their spectra. Consequently, precise identification of Herbig Ae/Be stars requires careful exclusion of contamination from other $H\alpha$-emitting sources.

\subsubsection{Removal of Galaxies, QSOs, CVs, and Algol Binaries}
The LAMOST DR7 spectroscopic dataset initially contained a non-negligible fraction of extragalactic sources (galaxies and QSOs), all identified via the``OBJTYP'' field in the header files. For the sake of caution, these were systematically removed during preliminary data cleaning. Nevertheless, since LAMOST only flagged galaxies and QSOs confirmed in the literature, we still cannot ensure that all of them have been completely removed. To address this, we first cross-matched our sample with published galaxy catalogues \citep{popescu1995search,ge2012double,comparat2016sdss}, but found no counterparts. We subsequently applied a redshift cut-off of $z\approx0.002$ (corresponding to a radial velocity of $\sim$600\,km\,s$^{-1}$) to exclude distant objects \citep{Sun2021}. Finally, we eliminated stellar sources identified as cataclysmic variables (CVs; \citealt{hou2020spectroscopically,Sun2021}) or Algol-type binaries \citep{budding2004catalogue} through additional catalogue matching.

\subsubsection{Removal of Late-Type Contaminants}\label{subsec:late_type}

A significant source of contamination in Herbig Ae/Be (HAeBe) star selection comes from late-type, low-mass pre-main-sequence (PMS) stars, particularly T-Tauri stars. Since HAeBes are intrinsically early-type stars, we systematically excluded late-type candidates (F, G, K, and M-type stars) from our sample. Following the methodology established by \citet{liu2015spectral}, who demonstrated the effectiveness of combining five spectral line indices (the equivalent widths of H$\gamma$, Mg, Fe, G-band, and TiO$_2$) for spectral-type classification in low-resolution LAMOST spectra, we applied these criteria to our candidate stars. Based on their empirical relations, we removed objects with G4300 $>1.5$\,\AA\ and  H$\gamma$ $<-4.0$\,\AA, ultimately retaining 54,609 early-type emission-line stars in our sample.

%\begin{figure}
%\begin{interactive}{js}{interactive.tar.gz}
%\plotone{Figure2.pdf}
%\caption{LAMOST J054113.78+273938.5. An Example of spectral fitting results %obtained by using the Pyhammer v2.0 program. The Blue spectrum represents one %spectrum from LAMOST DR7, while the red spectrum represents the best-fitting %temple. The top two panels present magnified regions around the $H\alpha$, Li I, %and Ca II triplet lines.}
%\label{f2}
%\end{figure}

%\subsection{Spectral classification and visually inspecting}
%\label{subsec:general}

To eliminate evolved stars and to better constrain the spectral types of the candidates, a spectral-type classification is required. To attain this objective, the automatic classifier MKCLASS \citep{Gray2014} was employed. When utilizing MKCLASS, we selected the spectral library \textit{libr18} and reduced the resolution from 2200 to 1800 to accommodate the LAMOST low-resolution spectra. After the spectra were automatically classified using MKCLASS, we conducted a visual inspection of the spectra. This step was taken to validate the outcomes of the automatic classification and to derive a more accurate result based on the available data. Through this verification process, we identified 5,129 early-type stars exhibiting unambiguous $H\alpha$ emission features, which were subsequently retained for further analysis.

%For this purpose, we employ PyHammer v2.0 \citep{roulston2020classifying}. Operating in the wavelength range 3650-10,200~\AA, PyHammer automatically compares key spectral lines with template spectra and determines the best-matching spectral type (from O to L), metallicity, and radial velocity (RV) for each star.
%
%First, the candidate spectra were processed using the MKCLASS algorithm. Following this automated classification, each spectrum underwent meticulous visual examination through comparison with both MKCLASS's early-type templates and confirmed HAeBe spectra from \citet{Zha2022}. Through this verification process, we identified 5,129 early-type stars exhibiting unambiguous $H\alpha$ emission features, which were subsequently retained for further analysis.

\subsection{Visual inspecting and SED fitting}
\label{subsec:sed}

When searching for emission line stars, the pollution of the emission nebulae is always a problem, since the $H\alpha$, N II, S II and O III emission lines of the ionized nebulae (planetary nebulae, supernova remnants, and HII regions) will overlap with stars of same sightline \citep{Zha2022}. Considering that the emission lines of the nebulae are usually accompanied with extended dust emission in the infrared, the near- to mid-infrared images of WISE are examined for each remaining candidates. 1835 stars showing notable infrared emission in all four bands are conserved, while stars showing ambiguous extended emission patterns are removed.

\begin{figure}
\plottwo{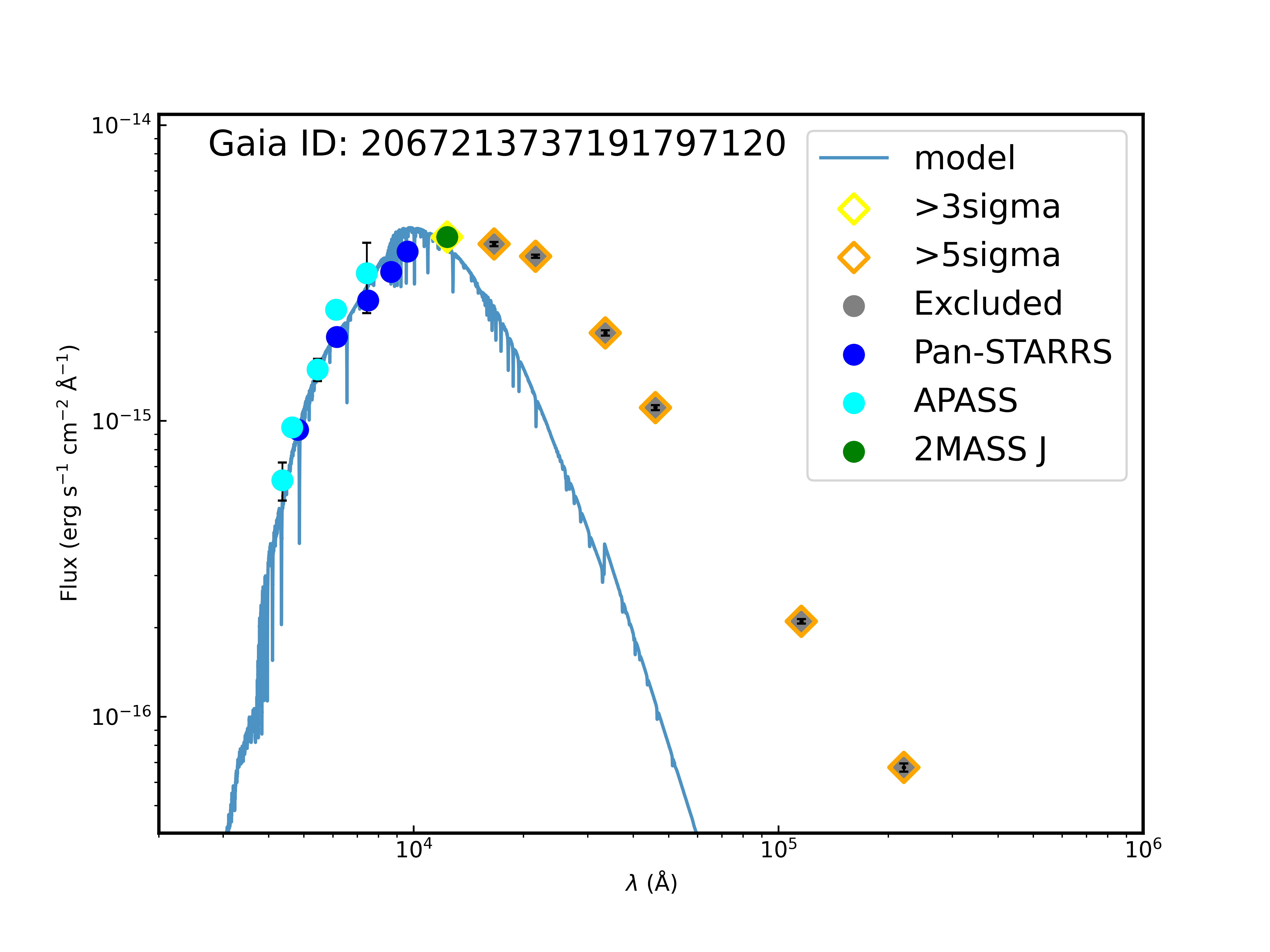}{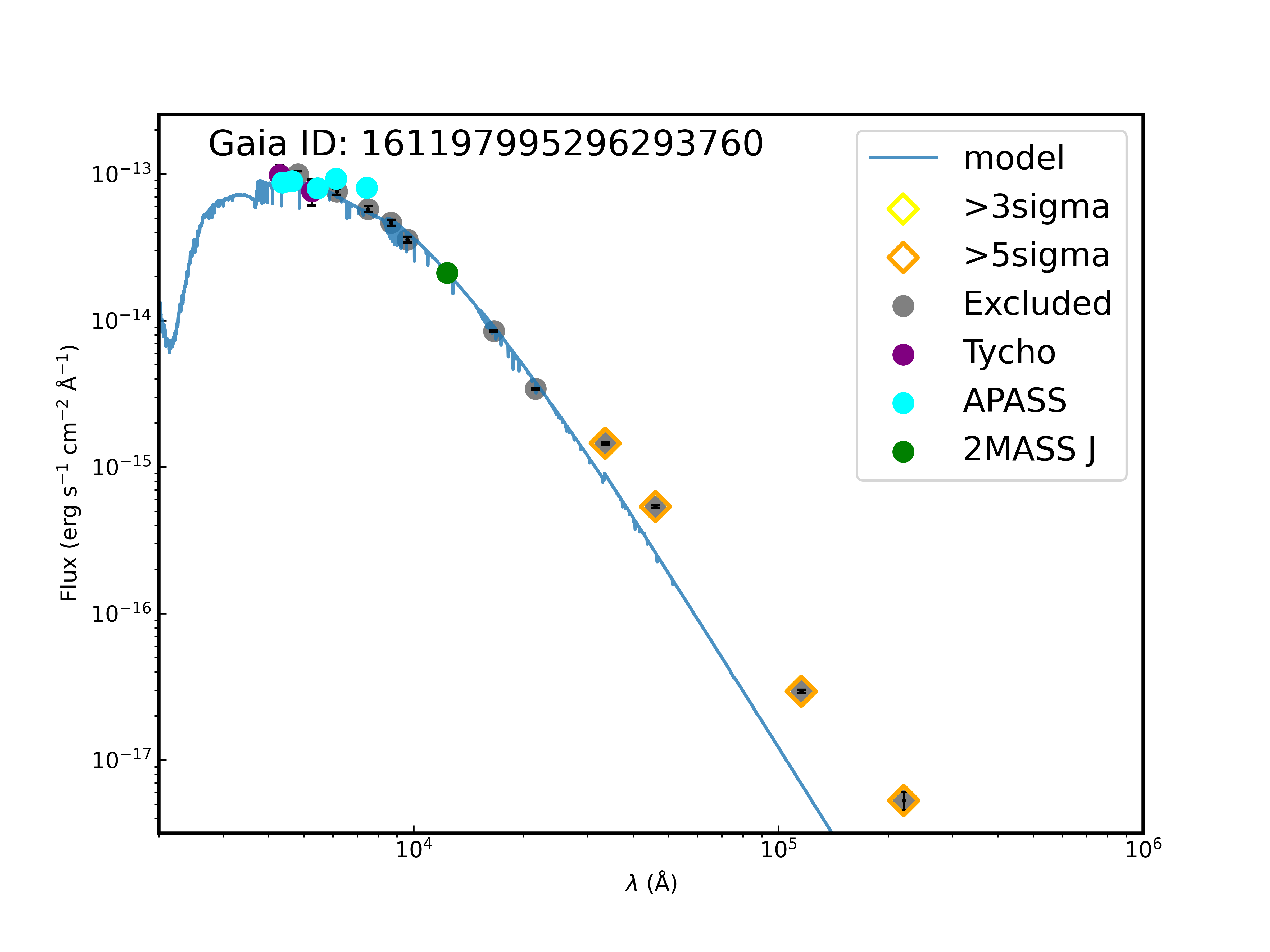}
\caption{Two SED fitting examples of the new identified Herbig Ae/Be star (the left panel) and candidate (right panel). The blue, cyan, purple and green dots are the available (unsaturated) photometric data of Pan-STARRS, APASS, Tycho-2 and the $J-$ band of 2MASS which are engaged in the SED fitting process, while gray solid dots denote the saturated bands or the infrared bands that are excluded from the SED fitting.  The blue line denotes the best fitted photospheric level of the model spectrum.
Infrared bands exhibiting significant excess emission are marked with yellow ($>3\sigma$) and orange ($>5\sigma$) diamonds.}
\label{SEDN}
\end{figure}

Following these steps, the late type stars, galaxies (including QSOs), CVs, Algol-type binaries and the evolved stars have been removed from the candidate list. But still there are some contaminations left, i.e., the CAeBes. As \cite{rivinius2013classical} argued in their work, that the infrared color excess of the CAeBes are weaker than HAeBes, especially in the mid-infrared. In light of this, the SEDs (Spectral Energy Distributions) of the candidates are examined to removed the contaminates of CAeBes.

In this work, the SED fitting is performed by minimizing the chi-square\,$\chi^{2}$ between the observed SED and the model SEDs. The photometric data used in the fitting include the $g_{\rm PS1}$, $r_{\rm PS1}$, $i_{\rm PS1}$, $z_{\rm PS1}$, $y_{\rm PS1}$ of Pan-STARRS (Panoramic Survey Telescope and Rapid Response System;  \citealt{Cha2016}), $B,V,g,r,i$ bands of the APASS9 (AAVSO Photometric All-Sky Survey; \citealt{Hen2016}) and $J-$ band of 2MASS (Two-Micron All Sky Survey; \citealt{Skr2006}). To account for potential saturation in bright stars, we also incorporate the $B_{\rm T}$ and $V_{\rm T}$ bands of Tycho-2 \citep{Hog2000} in case the Pan-STARRS and APASS optical bands are saturated.

The model SEDs are constructed through the following steps: 1) Convolving the BT-Settl model spectra \citep{Hau1999,Allard2012,All2014} with the Galactic average extinction curve ($R_{V}$=3.1) from \cite{wang2019optical} for varying extinction values. 2) Generating synthetic photometry for the Pan-STARRS, APASS, Tycho-2, and 2MASS bands. Since each BT-Settl template spectrum corresponds to a specific set of stellar parameters (i.e. effective temperature $T_{\rm eff}$, metallicity and surface gravity ${\rm log}\,g$), the optimal stellar parameters and the corresponding extinction value ($A_{V}$) can be simultaneously determined by identifying the minimum $\chi^{2}$ between the observed SED and the model SEDs. 
Using the spectral types of these candidates determined in the previous steps, we first estimate their effective temperatures $T_{\rm eff\_SpT}$ using the empirical relations from \cite{Pec2013} (for O9-A9) or \cite{Fang2017} (for F0-M9). To account for potential biases introduced by emission lines and accretion veiling in spectral typing \citep{Man2019}, we restrict the $\chi^{2}$ fitting to model templates within a defined temperature range around the spectroscopically derived $T_{\rm eff\_SpT}$. The size of this range is scaled based on $T_{\rm eff\_SpT}$ itself: $\pm$ 150\,K for $T_{\rm eff\_SpT} \leq$ 7000\,K, $\pm$ 300 K for 7000$ < T_{\rm eff\_SpT} \leq $12000\,K, $\pm$ 750 K for 12000$ < T_{\rm eff\_SpT}  \leq $20000\,K, and $\pm$  1500\,K for $T_{\rm eff\_SpT} > $20000\,K. In the SED fitting, the considered extinction range and the parameter space for stellar properties are outlined in Table \ref{tbl1}.

After applying the SED-fitting method to all 1835 HAeBe candidates, we find that 143 of them exhibit significant infrared excess. Among these, 92 stars are previously known HAeBes reported in the literature \citep{Hernandez_2005,Jose2012,Jose2013,Pan2020,Sko2020,guzman2021homogeneous,Grant_2022,vioque2022identification,Zha2022,Rom2023,Zhang2023}. Of the remaining 51 objects, 26 display pronounced IR excess (higher than 5$\sigma$ level) and are therefore considered newly identified HAeBes (see Table~\ref{tbln} and Figure~\ref{SEDAN} for details). The other 25 stars are currently classified as HAeBe candidates due to the moderate excess in the infrared (see Table \ref{tblc} and Figure \ref{SEDAC}). The H$\alpha$ emission line spectra of the 26 newly identified HAeBes and the 25 candidates are presented in Figure~\ref{NH} and Figure~\ref{CH}, respectively. In Figure \ref{SEDN}, we present the SED fitting results of a newly identified Herbig Ae/Be star (left panel) and a candidate star (right panel).

\begin{deluxetable*}{lcl}
\label{tbl1}
\tablecaption{Stellar Parameter Space and Sampling of the SED Fitting}
\tablehead{
\colhead{Stellar Parameters} & \colhead{Range} & \colhead{Step}}
\startdata
Extinction $A_{V}$ & 0.0--10.0\,mag & 0.1\,mag \\
Effective Temperature $T_{\rm eff}$ & 3000--35000\,K & Variable\tablenotemark{a} \\
Surface Gravity log\,$g$ & 2.0--5.0 & 0.5\,log\,$\rm  cm\,s^{-2}$ \\
Metallicity [Fe/H] & -0.5--0.5 & 0.25\,dex \\
Alpha Enhancement [$\alpha$/Fe] & 0.0 & Fixed \\
%Projected Rotational Velocity $v \sin i$ & 0\,km\,s$^{-1}$ & Fixed\tablenotemark{b}
\enddata
\tablenotetext{a}{The step for effective temperature is variable: 100\,K for $T_{\rm eff} \leq 7000$\,K, 200\,K for $7000 < T_{\rm eff} \leq 12000$\,K, 500\,K for $12000 < T_{\rm eff} \leq 20000$\,K, and 1000\,K for $20000 < T_{\rm eff} \leq 35000$\,K.}
%\tablenotetext{b}{The BT-Settl model templates assume $v \sin i = 0$\,km\,s$^{-1}$, as is standard for theoretical spectral libraries.}
\end{deluxetable*}

\section{Results}
\label{sec:result}

\subsection{Color-magnitude diagram}
\begin{figure}[htp]
\centering
\includegraphics[scale=0.56]{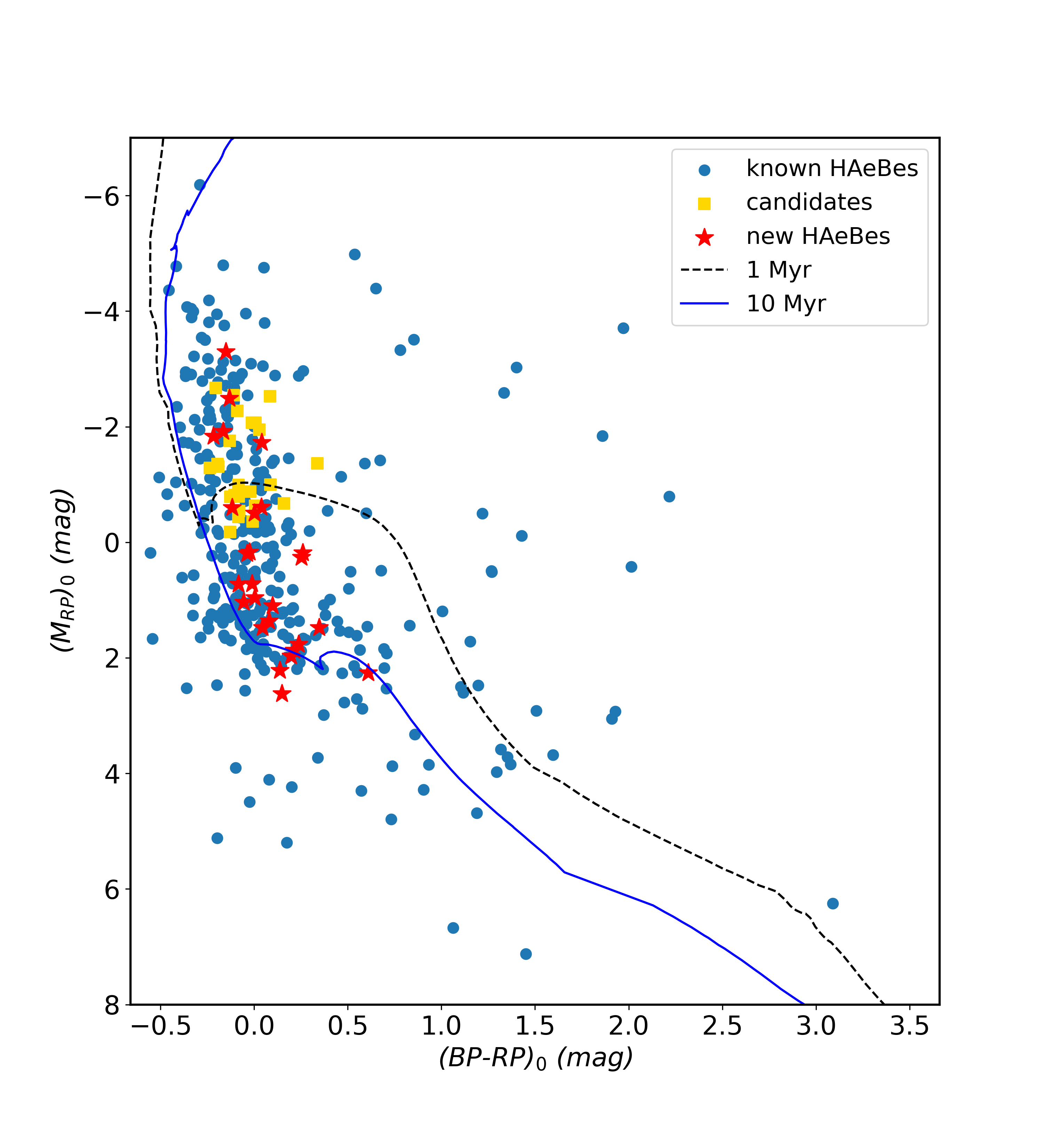}
\caption{The $M_{\rm RP}$ vs. $BP-RP$ color-magnitude diagram of the newly identified HAeBes (red asterisks) and the candidates (solid Wistia rectangles). The HAeBes of the literature are denoted as blue solid dots. Also plotted are the 1~Myr and 10~Myr isochrone of the PARSEC model for solar metallicity \citep{Bre2012}.}
\label{CMD}
\end{figure}

In Figure \ref{CMD}, we present the dereddened color-magnitude diagram (CMD) of the newly identified HAeBes in this study. For comparison, we also include previously known HAeBes from the literature (\citealt{Herbest_1999, Hernandez_2005, vioque2018gaia, Arun_2019, Wichittanakom_2020, guzman2021homogeneous, Grant_2022, vioque2022identification, Zha2022}). The interstellar extinction for these stars was remeasured using our SED fitting method (see Section \ref{subsec:sed}) and corrected for the corresponding photometric bands following the extinction law of \cite{wang2019optical}. The distances of the stars are adopted from \cite{Bai2021}.

\subsection{Age and Mass} \label{secAM}
HAeBes are intermediate-mass pre-main-sequence objects (2-8\,$M_{\odot}$), and their evolutionary stage and disk properties are critically relevant with their age and initial mass \citep{Mee2001}. Age estimates help constrain the timescales of disk dissipation and planet formation, while mass determines the radiative feedback and potential for triggering disk photoevaporation \citep{Alex2014}. In brief, younger HAeBe stars often exhibit stronger accretion and more prominent circumstellar disks, whereas older systems may transition into debris disks or naked photospheres \citep{Gorti2009}. To estimate stellar ages and masses, we constructed a CMD ($M_{\rm RP}$ vs. $BP-RP$) grid based on the PARSEC stellar evolutionary models.  We selected isochrones spanning 0.1-20\,Myr, with an age step of 0.1\,Myr, and linearly interpolated each isochrone into 1000 points to ensure high grid density. Each grid point corresponds to specific age and mass values.  By locating the HAeBe/candidate stars on this CMD, we derived their ages and masses by identifying the nearest model grid points.

\begin{figure}[htp]
\centering
\includegraphics[scale=0.56]{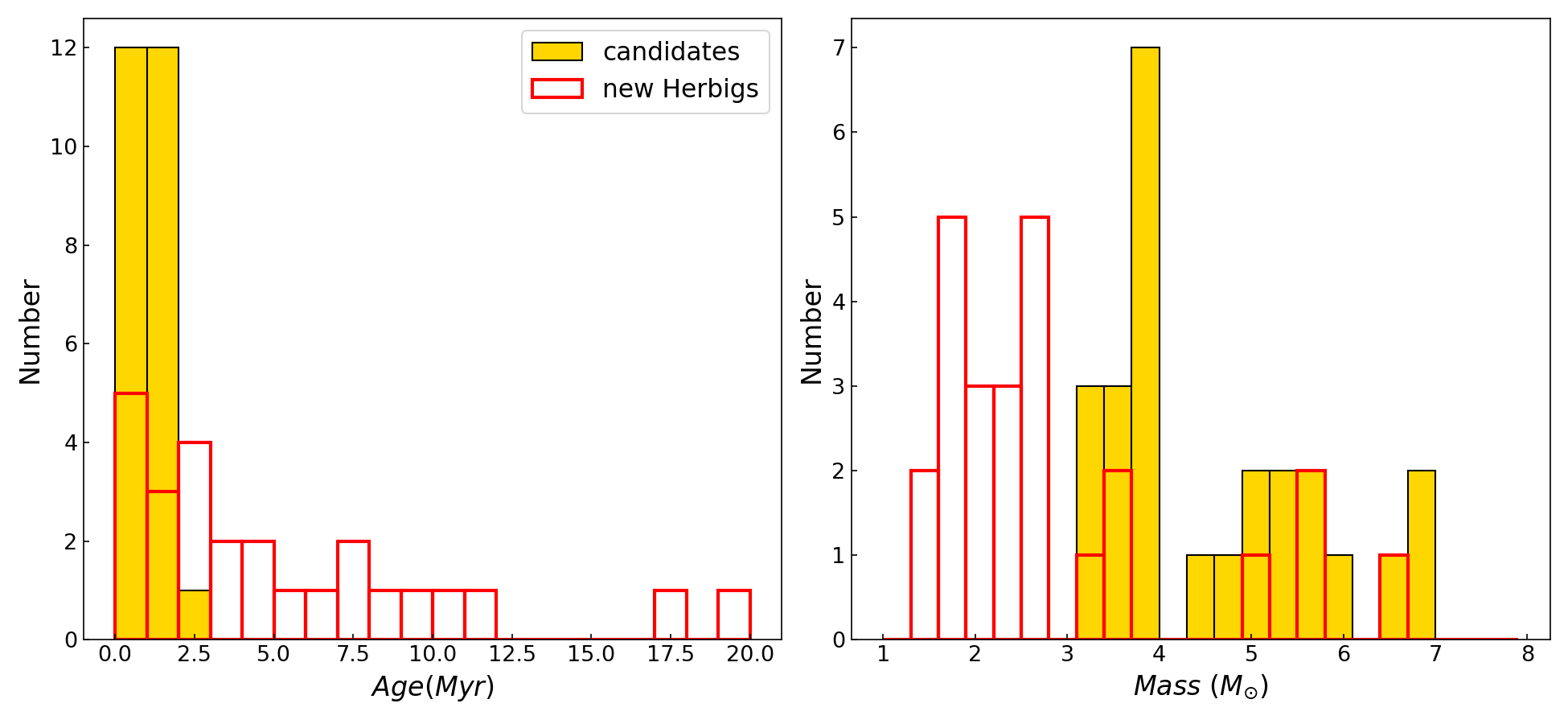}
\caption{The age (left panel) and mass (right panel) distribution of the newly identified HAeBes and the candidates. }
\label{AM}
\end{figure}

%\begin{figure}[htp]
%\centering
%\includegraphics[scale=0.6]{Mass.png}
%\caption{The Mass distribution of the new identified Herbig Ae/Be stars and the candidates.}
%\label{mass}
%\end{figure}

To ensure statistical robustness, we performed a 100 - fold Monte Carlo (MC) simulation for each star, incorporating photometric, extinction, and distance uncertainties. The median of these realizations was adopted as the final age and mass estimate, with the 1$\sigma$ standard deviation representing the measurement uncertainties (refer to Figure \ref{AM}, columns 6, 7, 8 and 9 of Table \ref{tbln} and Table \ref{tblc}). As depicted in the figure, the HAeBe candidates are relatively younger and more massive than the newly identified HAeBe stars. This is attributed to the fact that the HAeBe candidates are predominantly B-type stars, which typically have higher masses and undergo faster evolution. Consequently, their positions on the color-magnitude diagram (CMD) correspond to a younger age. It should be noted that HAeBes are variable stars, and factors such as the morphology and alignment of their circumstellar disks will influence their positions on the CMD. Moreover, HAe stars on the CMD occupy the transition zone between the pre-main sequence and main sequence phases, while HBe stars are located in the overlapping region where older B-type stars beginning their subgiant evolution intersect with young B-type stars.  Hence, we emphasize that age and mass estimates of HAeBe stars from their CMD positions require careful interpretation.

\subsection{Color-color diagrams}
\label{subsec:CCD}

%\begin{figure}
%\centering
%\plotone{Figure5.pdf}
%%\end{interactive}
%\caption{Color-color diagrams.
%%The left upper corner is the ((H-K$)_0$,(J-H$)_0$) diagram. And the right upper corner is the ((K-W1$)_0$, (H-K$)_0$) diagram. The left bottom corner is the ((K-W2$)_0$,(H-K$)_0$) diagram. And the right bottom corner is the ((K-W3$)_0$, (H-K$)_0$) diagram.
%In these four diagrams, CAeBes are grey points, known HAeBes are orange points, confirmed HAeBes are red and candidates are blue. HAeBes and candidates with an error of magnitudes \textgreater 0.05 mag are marked with triangle. Black dotted lines distinguish HAeBes from CBes.}
%\label{f5}
%\end{figure}

Due to the infrared radiation from the accretion disks of HAeBes, the infrared excess in near- and mid-infrared bands constitutes a characteristic feature of these objects. Notably, given that the infrared excess of CAeBes is predominantly concentrated in the near-infrared band, infrared color-color diagrams have proven to be an effective tool for distinguishing between CAeBes and HAeBes.

Based on the infrared color properties, \cite{finkenzeller1984herbig} proposed an empirical criterion of $(H-K)_0$ \textgreater 0.4 and $(K-L)_0$ \textgreater 0.8 to identify HAeBes. Subsequent studies by \cite{hou2016catalog} further refined this classification scheme through comprehensive analysis of infrared color-color diagrams. Their results demonstrated that HAeBes predominantly occupy the region defined by $(H-K)_{0}$\textgreater 0.4 and $(K-W1)_{0}$\textgreater 0.8, while CAeBes are primarily concentrated in the area characterized by $(H-K)_{0}$\textless 0.2 and $(K-W1)_{0}$\textless 0.5. The region between the HAeBes and CAeBes are regarded as the mixed region, where stars within this region might either be CAeBes or HAeBes \citep{Zha2022}.

\begin{figure}
\centering
\includegraphics[scale=0.56]{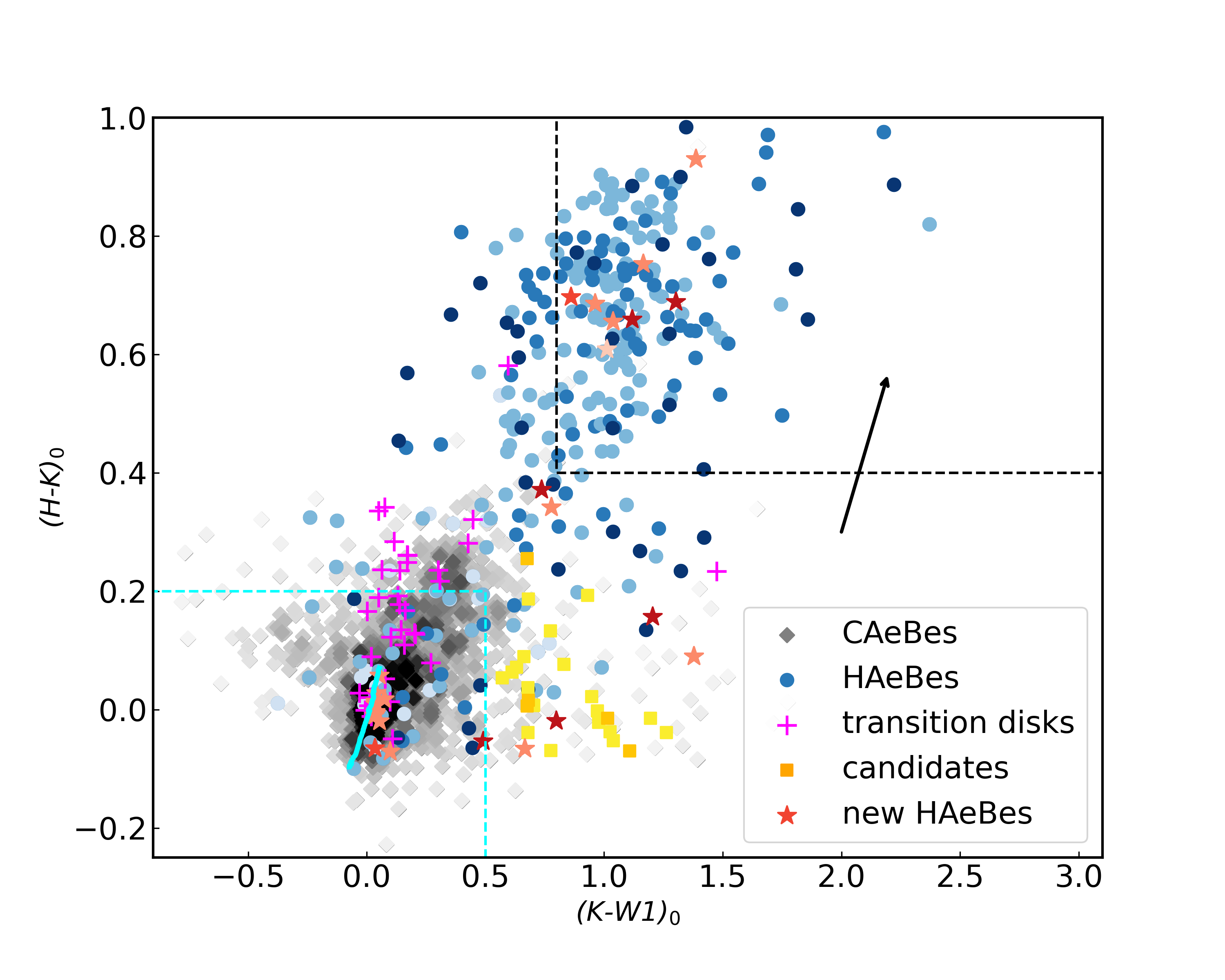}
%\end{interactive}
\caption{The $(K-W1)_0$ vs. $(H-K)_0$ color-color diagram of the previous known HAeBes (solid blue dots), newly identified HAeBes (red asterisks) and candidates (solid Wistia rectangles). The colors of the symbols represent the evolutionary stages of the protoplanetary disks, ranging from light to dark, which correspond to Class III, Class II, Flat-SED, and Class I. The previously identified CAeBes and PMS stars with transitional disks \citep{Cieza2012} are represented by gray diamonds (where color darkness indicates density) and magenta crosses, respectively. As a comparison, the regions of CAeBes and HAeBes that proposed by \cite{Zha2022} are denoted as cyan and black dashed lines, respectively. The cyan solid line indicates the intrinsic color indices of main-sequence B-type and A-type stars derived from the BT-settl stellar atmospheric model \citep{Pec2013}, while the black arrow shows the extinction vector of $A_{V}$=5.0.  }
\label{CCD1}
\end{figure}

Following their definition, we present the $(H-K)_{0}$ vs. $(K-W1)_{0}$ color-color diagram of the newly identified HAeBes of this work in Figure \ref{CCD1}. The identified CAeBes and HAeBes are also plotted as comparison. To better constrain the extinction correction of them, the extinction of CAeBes are also estimated with the SED fitting method and visually checked. As shown in the figure, approximately ten newly identified HAeBes are located in the HAeBe region, while five stars fall within the transition zone between the HAeBes and CAeBes. Notably, eleven stars along with dozens of previously confirmed HAeBes from the literature are situated in the CAeBe region. This is not unexpected, as their infrared excess is primarily observed in the mid-infrared bands ($W3$ and $W4$), with no significant excess detected in the near-infrared ($J$,$H$,$K$) or $W1$ bands (see Figure \ref{SEDAN}).

\subsection{Stages of the new Herbig Ae/Be stars}\label{stage}

%The infrared spectral index which refers to the slope of the SED between 2 and 20 $\mu m$:
%\begin{equation}
%\alpha=\frac{d\, \rm{log}(\lambda F_{\lambda})}{d\, \rm{log} \lambda},
%\end{equation}
%are introduced by \cite{Lada1984} to evaluate the evolutionary ``Stage'' (or ``Class'') of the YSOs, where $\lambda$ is the wavelength and $F_{\lambda}$ is the flux density at that wavelength. Referring to the evolutionary stages of the YSOs, except the deeply embedded Class 0 protostars (which are supposed to be undetectable), \cite{Lada1987} divided the infrared spectral index into three categories, i.e., Class I (stars with thick infalling envelope: $0<\alpha \lesssim3$), Class II (stars with accretion disk: $-2\lesssim \alpha \le0$) and Class III (stars with little to no circumstellar material: $-3<\alpha \lesssim-2$).
The infrared spectral index, which measures the slope of the SED between 2 and 20\,$\mu m$, is defined as:
\begin{equation}
\alpha=\frac{d\, \log(\lambda F_{\lambda})}{d\, \log \lambda},
\end{equation}
where $\lambda$ is the wavelength and $F_{\lambda}$ is the flux density at that wavelength. This index was introduced by \cite{Lada1984} to classify the evolutionary stage (or class) of young stellar objects (YSOs).

Based on the evolutionary stages of YSOs, \cite{Lada1987} categorized the infrared spectral index into three classes, excluding the deeply embedded Class 0 protostars (which are generally undetectable at the optical and infrared): YSOs with thick infalling envelopes (Class I, $0<\alpha \lesssim3$), YSOs with accretion disks (Class II, $-2\lesssim \alpha \leq0$) and YSOs with little to no circumstellar material (Class III, $-3<\alpha \lesssim-2$).

Later on, the infrared spectral index has been discussed across more infrared bands and more refined evolutionary stages of young stellar objects \citep{And1993,Greene1994,Evans2009}. Currently, the most widely adopted classification scheme is the four-stage system proposed by \cite{Greene1994}, defined as follows:
\begin{itemize}
\item Class I: $\alpha \geq 0.3$
\item Flat-SED: $-0.3 \leq \alpha < 0.3$
\item Class II: $-1.6 \leq \alpha < -0.3$
\item Class III: $\alpha < -1.6$.
\end{itemize}

In this study, we calculated the infrared spectral indices $\alpha$ for the newly identified HAeBes and candidates using photometric data of $K$, $W1$, $W2$, $W3$ and $W4$ bands. Following the classification criteria of \cite{Greene1994}, we categorized their evolutionary stages. Among the 26 newly identified HAeBes, we find six sources in the Class I Stage, two exhibiting flat SED, 17 classified as Class II, and only 1 object reaching the Class III stage. In contrast, the candidates shows a more evolved population distribution, dominated by 5 Class II and 20 Class III objects, suggesting these candidates represent more advanced evolutionary stages compared to the confirmed HAeBes (for more details, please refer to Figure \ref{CCD1}, column 10 of Table \ref{tbln} and Table \ref{tblc}).

\section{Discussion}
\label{sec:Dicussion}

%In this Section, our work is discussed. Some influence of WISE data and Ca emission line will be mentioned. We also discuss the process of examining our results, we cross-match our HAeBes catalogs with simbad database and Gaia x YSO database. Then, we discuss the color distribution of our stars.

In Section \ref{subsec:CCD}, we present the dereddened $(H-K)_{0}$ versus $(K-W1)_{0}$ color-color diagram for the newly identified HAeBes (Figure \ref{CCD1}). Intriguingly, approximately a dozen of these sources occupy the parameter space typically associated with CAeBes. The origin of this pattern lies in the fact that these stars exhibit significant mid-infrared excess in $W3$ and $W4$ bands, while showing little to no excess in the near-infrared bands ($J$, $H$, $K$, and $W1$ bands; see Figure \ref{SEDAN}  and \ref{SEDAC} for complete SED analysis).

This apparent classification discrepancy does not necessarily invalidate the Herbig Ae/Be nature of these stars. In reality, the evolution of pre-main-sequence (PMS) accretion disks represents a highly complex and not yet fully understood process. Multiple physical mechanisms, including viscous accretion, magnetohydrodynamic winds, photoevaporation, and stellar evolution effects, likely interact through non-linear coupling (see \citealt{Kom2025} for current perspectives on these uncertainties).

Particularly for intermediate- to high-mass PMS stars (i.e., HAeBes), the prevailing theoretical framework suggests that the strong UV/X-ray radiation from the central star drives inside-out disk dispersal, predominantly via photoevaporation processes \citep{Hol1994}. This model has gained substantial support from both observational studies and numerical simulations \citep{Mee2001,Gorti2009,Alex2014,Man2019,Her2024,Liu2024,Kom2025}. \cite{Alex2014} reviewed the theory and observation studies on how protoplanetary disks are dispersed, and they noticed that once photoevaporation becomes the major driver of disk evolution (mainly but not limited to class II/III phases), the disk will be rapidly cleared from the inside-out. This theory also gains support from the studies of the spectral energy distribution analyses. Notably, \cite{guzman2021homogeneous} conducted systematic SED fitting for 209 confirmed HAeBes using the Virtual Observatory SED Analyzer (VOSA). And their work suggests that the IR excess of HAeBes may begin from the $J$, $H$, or $K$ bands, or even longer wavelengths.

\begin{figure}
\centering
\includegraphics[scale=0.56]{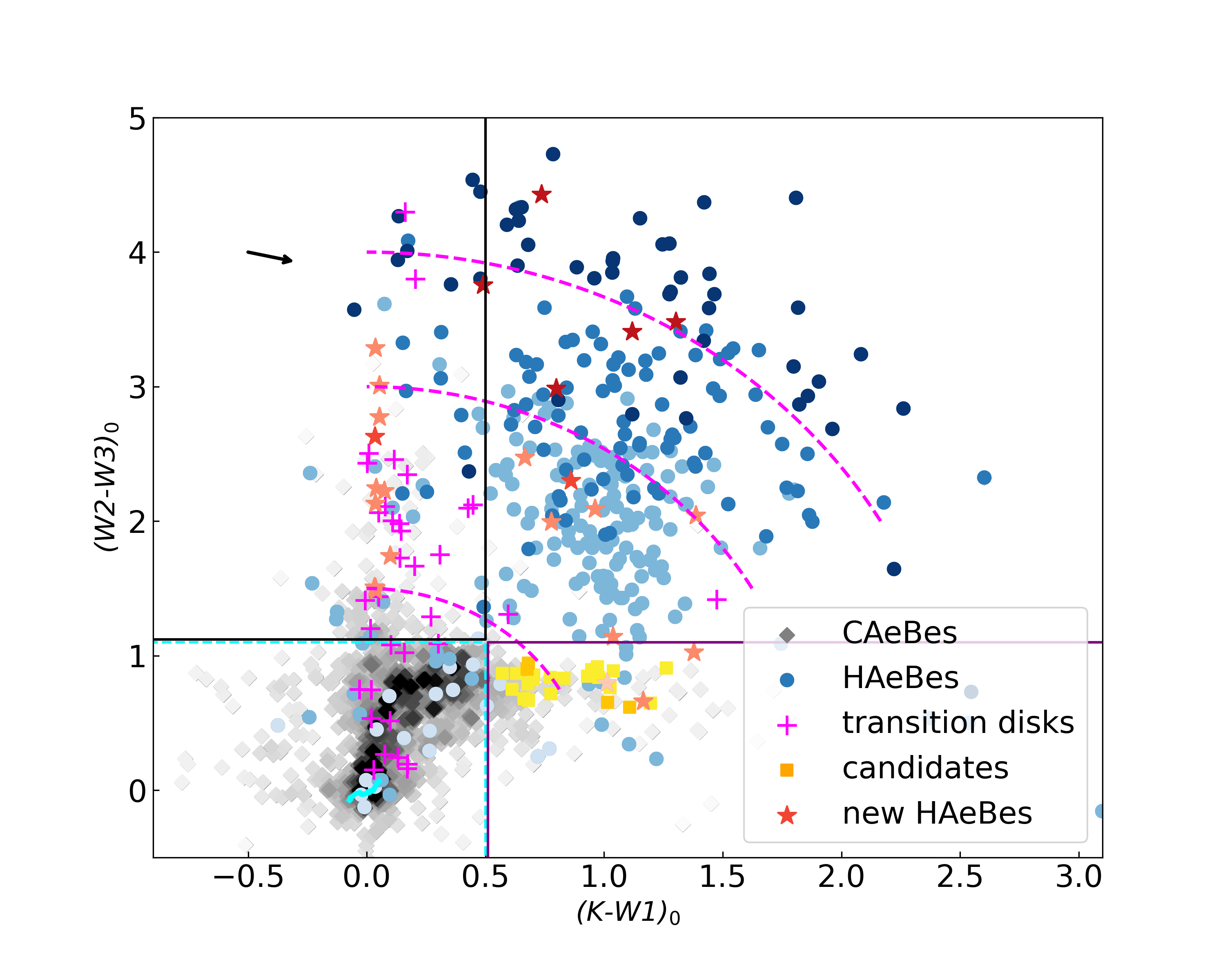}
%\end{interactive}
\caption{The $(K-W1)_0$ vs. $(W2-W3)_0$ color-color diagram of the previous known HAeBes (solid blue dots), newly identified HAeBes (red asterisks) and candidates (solid Wistia rectangles). The colors of the symbols represent the evolutionary stages of the protoplanetary disks, ranging from light to dark, which correspond to Class III, Class II, Flat-SED, and Class I. The previously identified CAeBes and PMS stars with transitional disks \citep{Cieza2012} are represented by gray diamonds (where color darkness indicates density) and magenta crosses, respectively. The cyan dashed line denotes the region of $(K-W1)_{0}$\textless 0.5 and $(W2-W3)_{0}$\textless 1.1 where most of the CAeBes lie at, while the black and purple solid lines denote the regions of stars with transition disks and globally depleted disks, respectively. The cyan solid line indicates the intrinsic color indices of main-sequence B-type and A-type stars derived from the BT-settl stellar atmospheric model \citep{Pec2013}, while the black arrow shows the extinction vector of $A_{V}$=5.0. The purple dashed lines represent three concentric ellipses centered on the intrinsic color indices of an A0 main-sequence stars (0,0), with an axis ratio of 1.6:1 and semi-major axes of 1.5, 3.0, and 4.0, respectively. These ellipses effectively distinguish HAeBes in evolutionary stages of Class III, Class II, Flat-SED, and Class I.}
\label{CCD2}
\end{figure}

Therefore, considering that some Herbig Ae/Be stars only exhibit IR excess in the mid- and far-infrared bands, incorporating mid-infrared photometric data into the construction of color-color diagrams will be a more reasonable solution. An effective classification scheme must satisfy three key criteria: (1) clear separation between HAeBes and CAeBes, (2) comprehensive coverage of HAeBes at different evolutionary stages, and (3) utilization of photometric bands with reliable observational data for most targets.

Based on these considerations, we systematically examined the infrared color indices of HAeBes and CAeBes. Our analysis reveals that CAeBes exhibit distinct clustering in two color indices $(K-W1)_0$ and $(W2-W3)_0$. In contrast, the known HAeBe stars predominantly occupy the remaining parameter space and are distinctly separated from the CAeBe population. This clear dichotomy is visually demonstrated in Figure \ref{CCD2}. Notably, when comparing the intrinsic color indices of the main-sequence stars, the distribution of CAeBe stars exhibits a clear reddening track and cutoffs in both color indices, as indicated by the high-density ``spine'' of the CAeBes. Therefore, we consider the region defined by
\begin{equation}
(K-W1)_0 \leq 0.5 \quad and \quad(W2-W3)_0 < 1.1
\end{equation}
as the distribution area for CAeBe stars.

\begin{figure}
\centering
\includegraphics[scale=0.86]{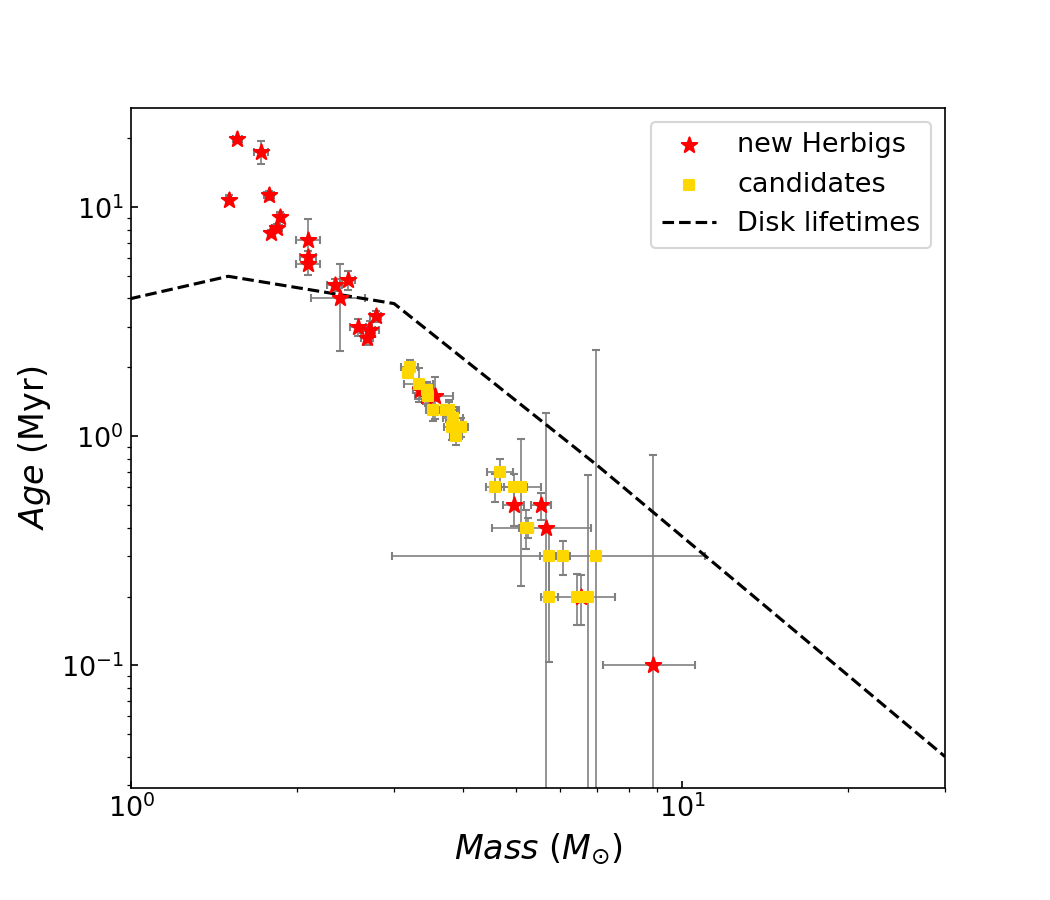}
%\end{interactive}
\caption{The mass and age distribution of the newly identified HAeBes (red asterisks) and the candidates (Wistia rectangles). The black dashed line indicates the survival timescales of circumstellar disks under the influence of viscous accretion and photoevaporation for stars of different masses \citep{Gorti2009}. }
\label{AvM}
\end{figure}

Another noteworthy phenomenon is the distinct pattern in the distribution of HAeBes on the $(K-W1)_0$ vs. $(W2-W3)_0$ color-color diagram. Overall, they are situated within a fan-shaped elliptical region centered on the intrinsic color indices of main-sequence stars, with an angle ranging from $\pi /6$ to $\pi /2$, and they exhibit a trend where the evolutionary stage becomes earlier as the distance from the center increases. This distribution aligns with the evolutionary patterns of HAeBes and also indicates that the $(K-W1)_0$ vs. $(W2-W3)_0$ effectively captures the evolutionary characteristics of HAeBes at various stages. To better show the distribution trend of HAeBes, we constructed three concentric ellipses centered at the point $(0,0)$ (the intrinsic color indices of an A0 main-sequence star), with the ratio of the semi-major axis $ a$ to the semi-minor axis $ b$ set at 1.6:1 and the semi-major axis lengths at 1.5, 3.0, and 4.0 (Figure \ref{CCD2}, magenta dashed lines). As shown in Figure \ref{CCD2}, Class III, Class II, and the Flat-SED HAeBes are predominantly located in the regions where $a<1.5$, $1.5\le a<3.0$, and $3.0\le a<4.0$, respectively. In contrast, the region where $a\ge4.0$ is mainly occupied by Class I stars. 

Based on observational data, the evolution of circumstellar disks around HAeBes can be divided into two categories: radially depleted disks (also known as transition disk) and globally depleted disks \citep{Fang2013}. The former exhibits little to no infrared excess in the near-infrared bands \citep{Ost1995,Mey2000,Mam2004,And2005,Car2005}, while the latter shows a uniform, weaker infrared excess extending up to 24 $\rm\mu m$ in the infrared bands \citep{Lada2006,Sic2008,Cur2009}. According to the distribution of HAeBes, the evolution of these two types of disks can be very clearly represented by two distinct regions on the $(K-W1)_0$ vs. $(W2-W3)_0$ color-color diagram. The first region is defined by $(K-W1)_0<$0.5 and $(W2-W3)_0>$1.1 (see the black line of Figure \ref{CCD2}). HAeBes in this region are mostly Class II and exhibit little to no infrared excess in the near-infrared bands but show significant infrared excess in the $W3$ band. Therefore, stars in this region are primarily identified as transition disk stars. The second region is defined by $(K-W1)_0>$0.5 and $(W2-W3)_0<$1.1 (purple line of Figure \ref{CCD2}). Almost all of the HAeBe candidates identified in this study are located in this region. Considering that these candidates are mainly Class III and exhibit uniformly weak excess in the infrared bands (see Figure \ref{SEDAC}), they are more consistent with the characteristics of globally evolved disks. To verify this, we conducted a preliminary analysis of these HAeBe candidates based on the survival timescales of accretion disks for stars of different masses. \cite{Gorti2009} made a time evolution of viscously accreting and photoevaporation on the circumstellar disks and concluded that protoplanetary disks around stars with $M_{*} \gtrsim3\,M_{\odot}$ have lifetimes shorter than 1\,Myr, and decreasing progressively with increasing stellar mass. While for stars of $M_{*} \lesssim3\,M_{\odot}$ the disk would typically survive for 3-6\,Myr (see the black dashed line of Figure \ref{AvM}). Though they claimed this relation is inconclusive for $M_{*} \lesssim3\,M_{\odot}$ stars, Figure \ref{AvM} clearly demonstrates that the circumstellar disks of HAeBe candidates (predominantly B-type stars) should be highly evolved and dissipated. 

We emphasize that the classification in this color-color diagram represents only a preliminary delineation based on the distributions of CAeBes and HAeBes. A more rigorous determination of the exact boundary regions would require both a larger sample size and precise measurements of stellar parameters (i.e., stellar mass, temperature and surface gravity) and extinction. We will address this issue in future work. Nevertheless, we contend that this classification offers a convenient and effective preliminary estimation of the evolutionary stages of HAeBe stars.

\section{SUMMARY}
\label{sec:Conclusion}

%In this work, using UMAP and SVM we seclected emission-line stars from the LAMOST DR7 database. Besides, screening out early-type stars and removed some errors of them. We then examined stars visually and the remain
In this work, using the LAMOST DR7 spectra and WISE, 2MASS photometric data, we identified 26 new Herbig Ae/Be stars and 25 candidates. The main result are summarized as follow:

$\bullet$ By applying UMAP and SVM to the low-resolution spectra of LAMOST DR7, we identified $\sim$240 thousands potential H$\alpha$ emission objects. Then we removed the contamination of galaxies (including QSOs), CVs, Algol-type binaries, evolved stars and emission line nebulae. With the criteria of G4300 $>$ 1.5\,\AA\ and H$\gamma$ $<$ -4.0\,\AA  \,\,and spectral classification method, the potential late-type contaminates are removed.

$\bullet$ We verified the 2MASS and WISE images of the candidates and used SED fitting to search for stars with significant IR excess. As a result, we identified 26 new Herbig Ae/Bes that showing obvious IR excess and 25 candidates with modest excess in the infrared.

$\bullet$  We find that Herbig Ae/Be stars are primarily distributed in a fan-shaped elliptical region on the $(K-W1)_0$ vs. $(W2-W3)_0$ color-color diagram, centered on the intrinsic color indices of main-sequence stars, with a semi-major axis $a$ to semi-minor axis $b$ ratio of 1.6:1 and an angle ranging from $\pi/6$ to $\pi/2$. A clear evolutionary gradient is observed for Herbig Ae/Be stars on this diagram, with those in the Class III, Class II, Flat-SED, and Class I evolutionary stages being effectively distinguished by concentric ellipses with semi-major axes of $a$=1.5, $a$=3.0, and $a$=4.0.

$\bullet$ The $(K-W1)_0$ vs. $(W2-W3)_0$ color-color diagram can also serve as an effective way to classify the radially depleted disks and globally depleted disks. The first category is mainly located at $(K-W1)_0 < 0.5$ and $(W2-W3)_0 > 1.1$, while the latter primarily concentrates at $(K-W1)_0 > 0.5$ and $(W2-W3)_0 < 1.1$.

\section{ACKNOWLEDGEMENTS}
{We thank the anonymous referee for the thorough reviews of our manuscript and constructive advice. W. Y. C. acknowledges the support from the National Key R\&D Program of China with grant No. 2024YFA1611903. M. F. acknowledges the support from the National Key R\&D Program of China with grant No. 2023YFA1608000. This work is supported by the National Natural Science Foundation of China (NSFC) under grants No. 12173013 and No. 12003045, the Science Foundation of Hebei Normal University under grant No. L2022B07, the project of Hebei Provincial Department of Science and Technology under grant No. 226Z7604G, the Hebei NSF under grant No. A2023205036, and the Key Development Foundation of Hebei Normal University under grant No. L2025ZD01.
%J.M.L. acknowledges the support from the National Natural Science Foundation of China (NSFC) with grant 12003045, the Science Foundation of Hebei Normal University with grant L2022B07 and the Hebei NSF with grant A2023205036.
The Guoshoujing Telescope (the Large Sky Area Multi-Object Fiber Spectroscopic Telescope, LAMOST) is a National Major Scientific Project built by the Chinese Academy of Sciences. Funding for the project has been provided by the National Development and Reform Commission. LAMOST is operated and managed by the National Astronomical Observatories, Chinese Academy of Sciences. This work has made use of data from the European Space Agency (ESA) mission Gaia (https://www.cosmos.esa.int/gaia), processed by the Gaia Data Processing and Analysis Consortium (DPAC, https://www.cosmos.esa.int/web/gaia/dpac/consortium). Funding for the DPAC has been provided
by national institutions, in particular the institutions participating in the Gaia Multilateral Agreement. Substantial data processing in this work was executed through the TOPCAT software \citep{Taylor2005}.}

\bibliography{reference}{}
\bibliographystyle{aasjournal}

\appendix
%\section{Source information}

\begin{figure}
\plotone{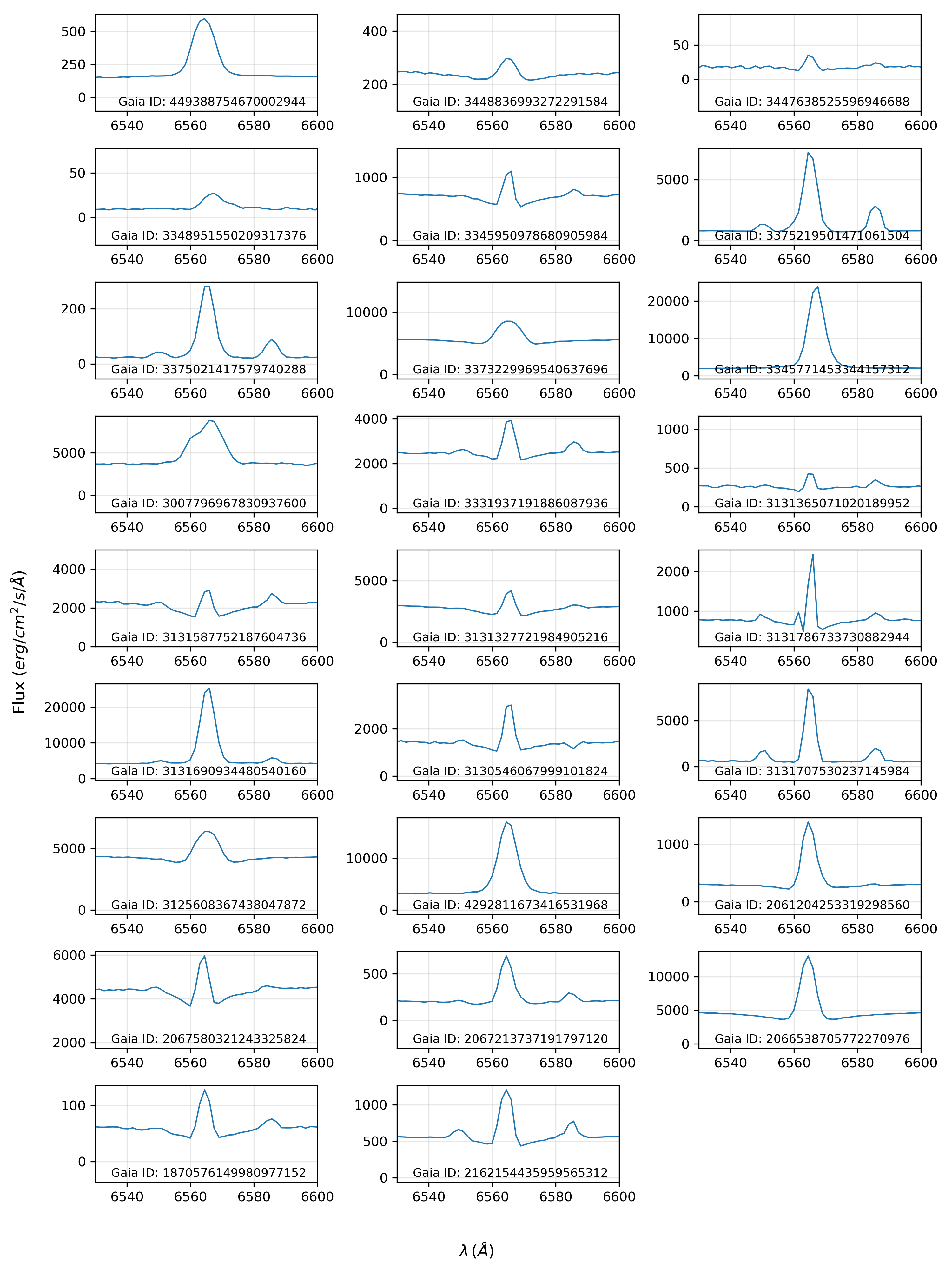}
\caption{The H$\alpha$ emission of the newly identified Herbig Ae/Be stars in this work.}
\label{NH}
\end{figure}

\begin{figure}
\plotone{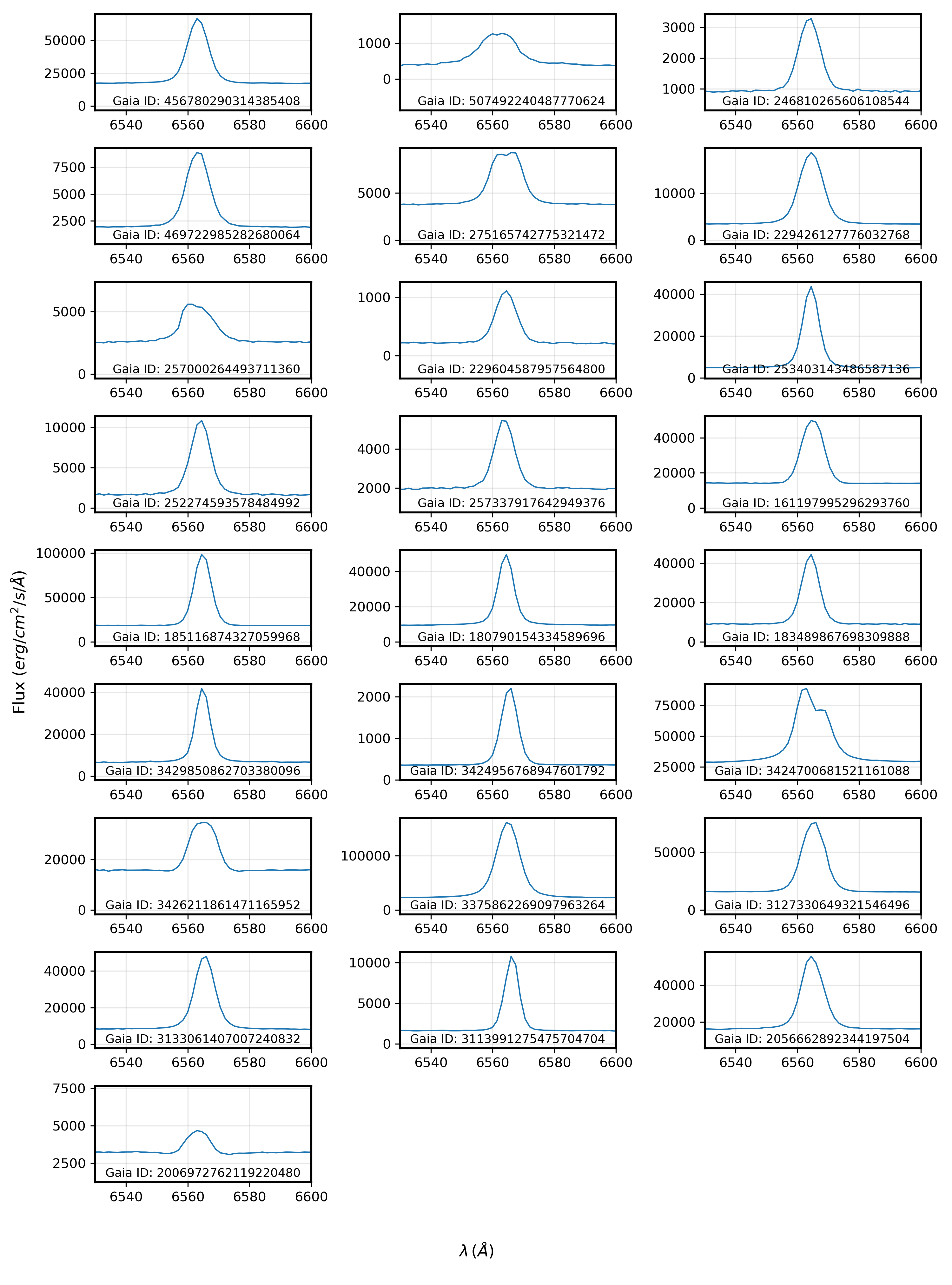}
\caption{The H$\alpha$ emission of the HAeBe candidates.}
\label{CH}
\end{figure}

\begin{figure}
\plotone{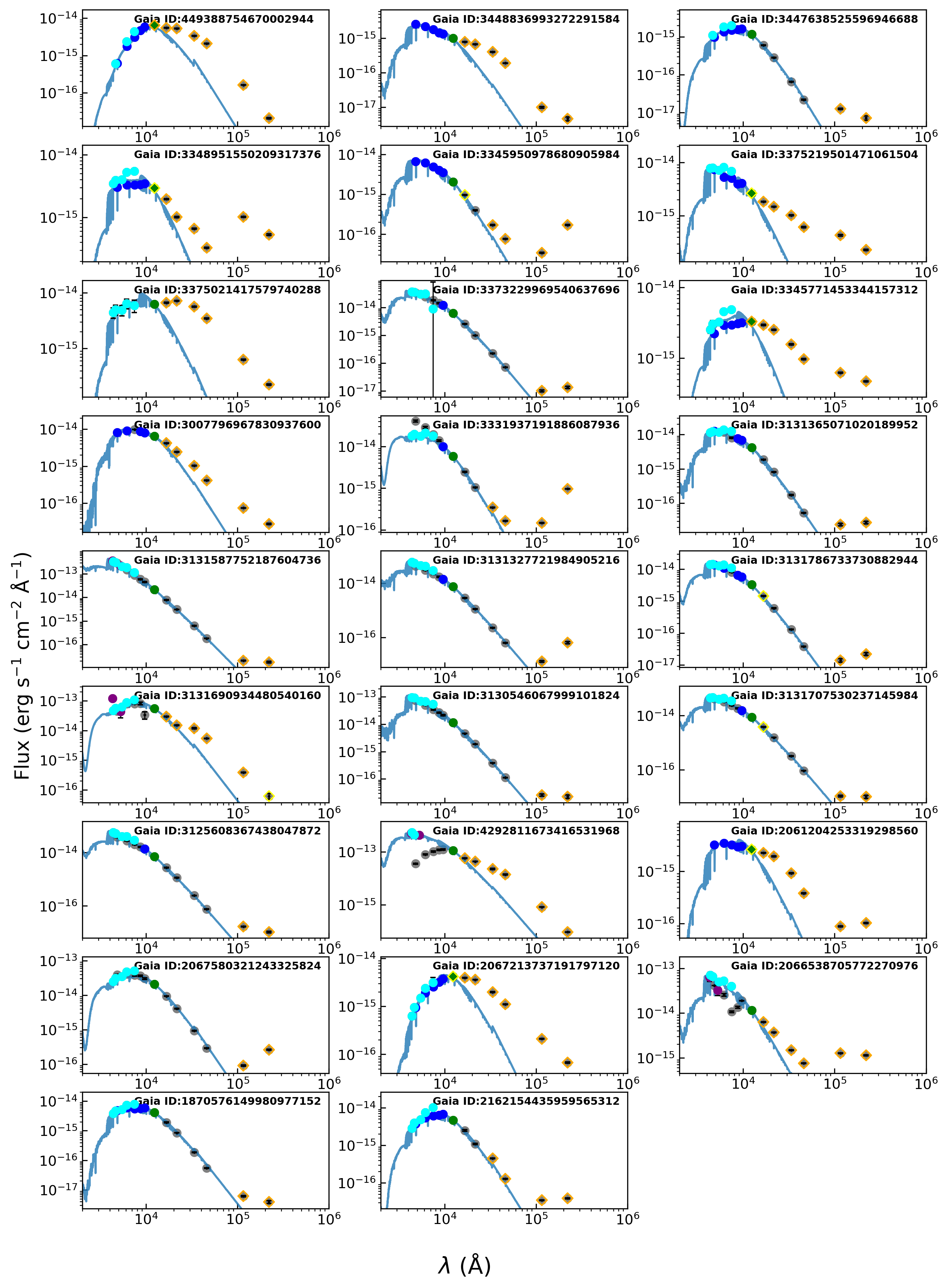}
\caption{SED fitting result for all the newly identified Herbig Ae/Be stars. The blue, cyan, purple and  green dots are the available (unsaturated) photometric data of Pan-STARRS, APASS, Tycho-2 and the $J-$ band of 2MASS which are engaged in the SED fitting process, while gray solid dots denote the saturated bands or the infrared bands that are excluded from the SED fitting.  The blue line denotes the best fitted photospheric level of the model spectrum.
Infrared bands exhibiting significant excess emission are marked with yellow ($>3\sigma$) and orange ($>5\sigma$) diamonds.}
\label{SEDAN}
\end{figure}

\begin{figure}
\plotone{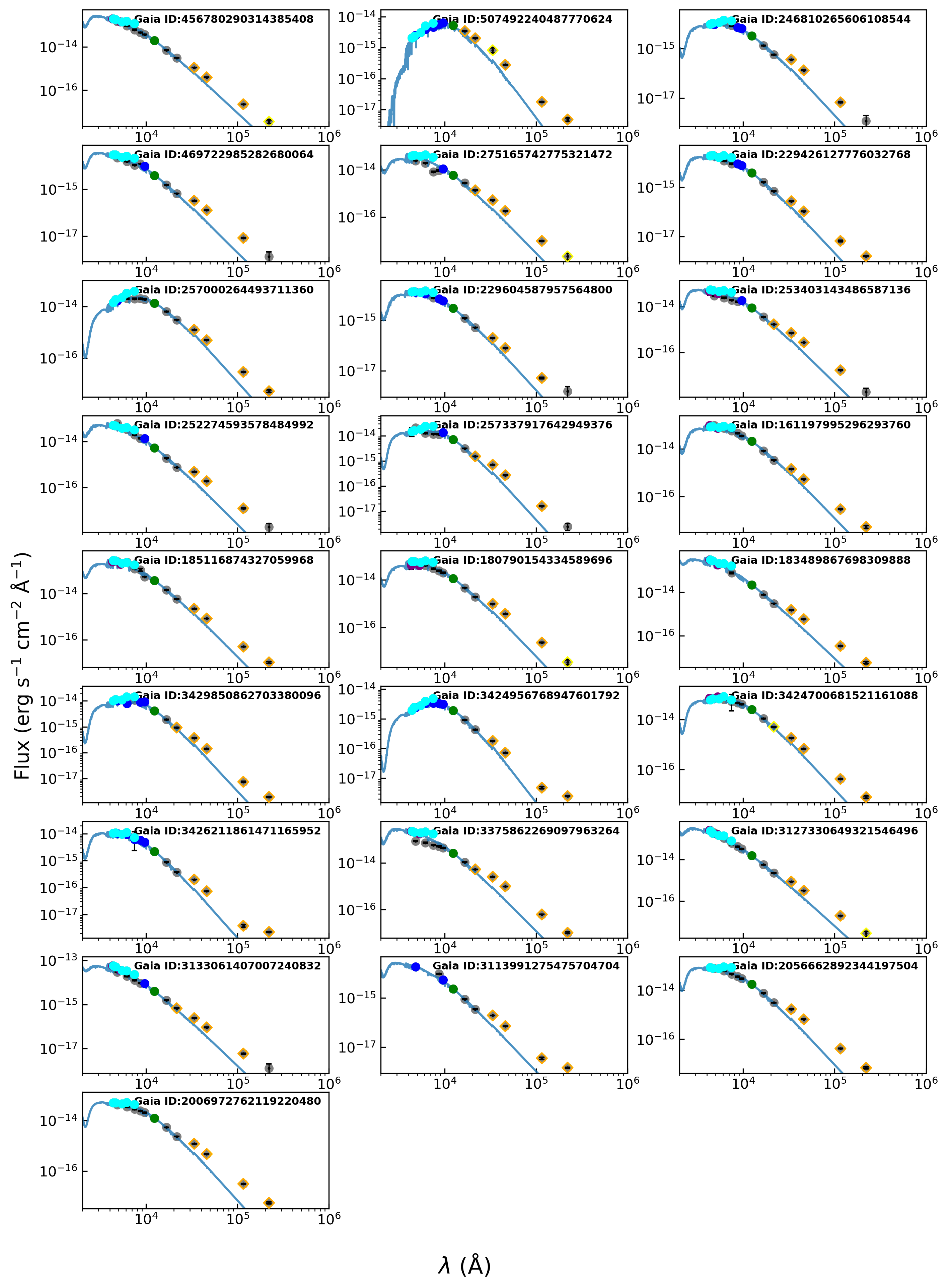}
\caption{Same as Figure \ref{SEDAN}, but for the 25 HAeBe candidates.}
\label{SEDAC}
\end{figure}

\begin{sidewaystable}[htbp]
\caption{The newly identified Herbig Ae/Be stars of this work.}
\label{tbln}
\begin{tabular}{lrrlllllll}
\hline\hline
Gaia ID & RAJ2000 & DEJ2000 & SpT &  $A_{V}$ & Age\textsuperscript{1} & e\_Age & $M_{*}$\textsuperscript{1}  & $e\_M_{*}$ & Stage\textsuperscript{2} \\
\hline
&(deg)&(deg)& &(mag) & (Myr)&(Myr)& ($M_{\odot}$)&($M_{\odot}$)& \\
\hline
449388754670002944 & 52.281420 & 57.026023 & B5 & 6.50 & 0.5 & 0.08 & 4.9 & 0.22 & Class II \\
3448836993272291584 & 84.155662 & 33.001109 & A5 & 2.00 & 17.5 & 1.97 & 1.7 & 0.05 & Class II \\
3447638525596946688 & 84.418818 & 31.169562 & B8 & 3.90 & 7.1 & 1.76 & 2.1 & 0.10 & Class II \\
3348951550209317376 & 89.565613 & 16.527038 & A1 & 3.10 & 6.0 & 0.38 & 2.1 & 0.06 & Class I \\
3345950978680905984 & 92.254985 & 15.719783 & A3 & 1.70 & 8.1 & 0.29 & 1.8 & 0.03 & Class I \\
3375219501471061504 & 92.261559 & 20.611910 & A1 & 2.10 & 5.6 & 0.55 & 2.1 & 0.10 & Class I \\
3375021417579740288 & 92.723581 & 20.539792 & A3 & 3.60 & 4.1 & 1.76 & 2.4 & 0.27 & Class II \\
3373229969540637696 & 92.855687 & 17.504833 & B8 & 1.60 & 2.9 & 0.16 & 2.7 & 0.07 & Class II \\
3345771453344157312 & 93.302890 & 15.343515 & B7 & 3.90 & 1.5 & 0.30 & 3.6 & 0.28 & Class I \\
3007796967830937600 & 93.666489 & -6.488165 & F8 & 1.80 & 10.7 & 0.50 & 1.5 & 0.02 & Class II \\
3331937191886087936 & 93.756571 & 12.332016 & B1 & 2.80 & 0.4 & 0.73 & 5.6 & 1.07 & Class I \\
3131365071020189952 & 97.198241 & 4.912075 & A7 & 1.50 & 7.7 & 0.25 & 1.8 & 0.02 & Class II \\
3131587752187604736 & 97.472540 & 5.225091 & B9 & 0.50 & 3.3 & 0.14 & 2.8 & 0.04 & Class II \\
3131327721984905216 & 97.827064 & 4.850411 & A5 & 0.60 & 9.1 & 0.45 & 1.9 & 0.03 & Class II \\
3131786733730882944 & 97.858365 & 5.474974 & B8 & 1.90 & 2.9 & 0.28 & 2.7 & 0.11 & Class II \\
3131690934480540160 & 98.406235 & 4.813071 & B1 & 4.00 & 0.1 & 0.77 & 8.9 & 1.53 & Class II \\
3130546067999101824 & 98.556884 & 4.542677 & A7 & 0.60 & 11.3 & 0.50 & 1.8 & 0.04 & Class II \\
3131707530237145984 & 98.585339 & 5.169352 & A1 & 1.30 & 19.9 & 0.00 & 1.6 & 0.03 & Class II \\
3125608367438047872 & 101.138408 & 0.693570 & B9 & 1.20 & 1.6 & 0.14 & 3.3 & 0.10 & Class II \\
4292811673416531968 & 290.698150 & 4.582640 & B7 & 2.10 & 0.2 & 0.00 & 6.6 & 0.15 & Class III \\
2061204253319298560 & 305.494893 & 38.931951 & A1 & 3.20 & 4.8 & 0.58 & 2.5 & 0.08 & SED Flat \\
2067580321243325824 & 306.410170 & 40.432314 & B2 & 3.70 & 0.4 & 0.05 & 5.6 & 0.22 & SED Flat \\
2067213737191797120 & 306.729093 & 39.629405 & A7 & 4.60 & 3.0 & 0.26 & 2.6 & 0.08 & Class II \\
2066538705772270976 & 310.278283 & 42.519901 & A0 & 1.50 & 4.6 & 0.27 & 2.3 & 0.06 & Class I \\
1870576149980977152 & 310.739688 & 36.318630 & A5 & 2.80 & 2.7 & 0.18 & 2.7 & 0.07 & Class II \\
2162154435959565312 & 314.905652 & 43.518192 & B8 & 4.00 & 1.5 & 0.10 & 3.4 & 0.08 & Class II \\
\hline
\end{tabular}
%\begin{tablenotes}
%\scriptsize{
%\item[1] \hspace{3cm}\textsuperscript{1}The age and mass of the stars are determined based on the grid points of stellar evolutionary models that best match their \\ \hspace{13cm} positions on the CMD (for more details please refer to Secrion \ref{AM}). Caution should be exercised when using these values.
%\item[2] \hspace{3cm}\textsuperscript{2}The evolutionary stage of stars that derived by the infrared spectral index $\rm \alpha$ (refer to Section \ref{stage}).}
%\end{tablenotes}
\tablecomments{
\textsuperscript{1}The age and mass of the stars are determined based on the grid points of stellar evolutionary models that best match their positions on the CMD (for more details please refer to Section \ref{secAM}). Caution should be exercised when using these values.
\textsuperscript{2}The evolutionary stage of stars that derived by the infrared spectral index $\rm \alpha$ (refer to Section \ref{stage}).
}
\end{sidewaystable}

\begin{sidewaystable}[htbp]
\caption{The Herbig Ae/Be candidates of this work.}
\label{tblc}
\small
\begin{tabular}{lrrlllllll}
\hline\hline
Gaia ID & RAJ2000 & DEJ2000 & SpT &  $A_{V}$ & Age\textsuperscript{1}  & e\_Age & $M_{*}$\textsuperscript{1}  & $e\_M_{*}$ & Stage\textsuperscript{2} \\
\hline
&(deg)&(deg)& &(mag) & (Myr)&(Myr)& ($M_{\odot}$)&($M_{\odot}$)& \\
\hline
456780290314385408 & 33.873105 & 55.962311 & B2 & 1.50 & 0.4 & 0.05 & 5.2 & 0.15 & Class III \\
507492240487770624 & 34.700102 & 60.920984 & F0 & 3.70 & 0.6 & 0.08 & 4.6 & 0.17 & Class II \\
246810265606108544 & 59.968010 & 48.227751 & B3 & 2.60 & 1.1 & 0.12 & 3.8 & 0.13 & Class III \\
469722985282680064 & 60.242131 & 57.379914 & B2 & 1.80 & 1.2 & 0.11 & 3.8 & 0.13 & Class III \\
275165742775321472 & 62.835946 & 53.234534 & B7 & 1.50 & 1.2 & 0.12 & 3.5 & 0.10 & Class III \\
229426127776032768 & 64.171469 & 43.073661 & B5 & 2.00 & 1.0 & 0.09 & 4.0 & 0.11 & Class III \\
257000264493711360 & 65.846811 & 46.144163 & B3 & 3.70 & 0.2 & 0.00 & 6.4 & 0.12 & Class III \\
229604587957564800 & 66.086165 & 44.250515 & B7 & 2.00 & 1.5 & 0.13 & 3.5 & 0.11 & Class II \\
253403143486587136 & 66.592663 & 44.008730 & B2 & 2.10 & 0.6 & 0.09 & 4.9 & 0.28 & Class III \\
252274593578484992 & 67.502742 & 42.383216 & B5 & 1.30 & 1.0 & 0.08 & 3.9 & 0.10 & Class III \\
257337917642949376 & 69.209208 & 47.688809 & B1 & 3.20 & 0.2 & 0.02 & 6.1 & 0.18 & Class III \\
161197995296293760 & 72.662572 & 32.721869 & B5 & 2.20 & 1.6 & 0.11 & 3.4 & 0.08 & Class III \\
185116874327059968 & 76.565628 & 34.952663 & B5 & 1.60 & 1.2 & 0.07 & 3.7 & 0.08 & Class III \\
180790154334589696 & 80.080003 & 32.138043 & B5 & 2.00 & 0.2 & 0.09 & 5.8 & 0.21 & Class III \\
183489867698309888 & 83.006894 & 36.588954 & B5 & 0.90 & 1.9 & 0.10 & 3.2 & 0.06 & Class III \\
3429850862703380096 & 88.908436 & 25.859303 & B2 & 3.10 & 0.7 & 0.09 & 4.7 & 0.26 & Class III \\
3424956768947601792 & 89.923502 & 24.070006 & B2 & 3.80 & 1.7 & 0.28 & 3.3 & 0.20 & Class II \\
3424700681521161088 & 90.764154 & 23.793712 & B2 & 2.90 & 0.2 & 0.40 & 6.7 & 0.92 & Class III \\
3426211861471165952 & 91.061535 & 24.067333 & B2 & 2.30 & 0.6 & 0.48 & 5.2 & 0.55 & Class II \\
3375862269097963264 & 94.944194 & 21.151134 & B2 & 1.60 & 0.4 & 0.03 & 5.3 & 0.10 & Class III \\
3127330649321546496 & 102.118131 & 3.132206 & B5 & 0.90 & 2.0 & 0.16 & 3.2 & 0.10 & Class III \\
3133061407007240832 & 103.028163 & 7.392794 & B5 & 0.90 & 1.2 & 0.15 & 3.8 & 0.15 & Class III \\
3113991275475704704 & 103.392069 & 1.102357 & B2 & 1.60 & 1.2 & 0.14 & 3.8 & 0.16 & Class II \\
2056662892344197504 & 308.943751 & 35.788918 & B2 & 2.20 & 0.2 & 0.08 & 5.7 & 0.20 & Class III \\
2006972762119220480 & 341.316490 & 56.627713 & B1 & 2.50 & 0.2 & 1.59 & 7.1 & 3.73 & Class III \\
\hline
\end{tabular}
\tablecomments{
\textsuperscript{1}The age and mass of the stars are determined based on the grid points of stellar evolutionary models that best match their positions on the CMD (for more details please refer to Section \ref{secAM}). Caution should be exercised when using these values.
\textsuperscript{2}The evolutionary stage of stars that derived by the infrared spectral index $\rm \alpha$ (refer to Section \ref{stage}).
}
\end{sidewaystable}


\end{CJK*}
\end{document}